*Original Article*

# Transforming Tarlac State University (TSU) Gymnasium to a Nearly Zero-Energy Building through Integration of a Solar Photovoltaic (PV) System

Rafael R. Yumul[1], Enalyn T. Domingo[2]

[1,2]Electrical and Electronics Engineering Department, Tarlac State University, Tarlac, Philippines.

[2]Corresponding Author : enalyntiamzondomingo@gmail.com

***Abstract -*** *The study is anchored to the principles of Nearly-Zero Energy Building (NZEB).  It aimed to transform the Tarlac State University Gymnasium into a facility with energy-efficient equipment to contribute to reducing carbon footprints by integrating a solar PV system as its renewable energy source.  The researchers found out that the electrical infrastructure of the Gym was outdated, and the lighting was not energy efficient, and there were too few convenience or power outlets. There was also insufficient cooling equipment to maintain a comfortable temperature.  Analysis shows that the payback period is within the average range, making it a cost-effective investment for the University.  Aside from the cost of the PV System, adherence to engineering design standards will mean additional costs to replace the metal halides with LED high bay lamps, installation of additional air conditioning units, and provision of additional convenience outlets. These additional costs should be considered when evaluating the feasibility of the project.  It is recommended that the integrity of the existing roof system of the Gymnasium be considered.  The total cost of putting up the whole electrical system, including new lighting, cooling, and convenience loads, must be calculated to determine the total cost of implementing the whole NZEB project. Other factors in the economic evaluation may be considered to determine a more stringent result.*

## 1. Introduction
There is a rise in the Earth's average temperature, and if the rise is steady, A global temperature of 1.5 °C will be reached between 20230 and 2052.  The impact of the rise includes global changes in climate, resulting in severe droughts and heavy precipitation in different areas, thereby affecting food security and sea levels [1].  PAGASA has begun to observe the same event in the climate system since the mid-20th century, which, according to the agency, is due to the observed increase in Greenhouse Gas (GHG) concentrations from human activities. As such, country leaders are actively seeking solutions to address these impacts. In 2022, the United Nations' Environment Programme (UNEP) convened the UN Climate Change Conference, which was attended by world leaders. The report highlights funding allocations to support several climate change adaptation programs.  These initiatives focus on protecting ecosystems like coral reefs in the Philippines and riverbanks in Fiji. They also aim to establish alternative power sources, such as a hydroelectric plant in Zimbabwe, and improve energy efficiency with new regulations in Pakistan and Indonesia [2]. Future risks related to climate issues would be reduced through adaptation and mitigation programs [3].

Countries like the Philippines also actively participate in global efforts to combat the effects of global warming. The commitment outlined in R.A. 9513 (Renewable Energy Act of 2008) underscores the government's dedication to utilizing renewable energy resources [4]. The Philippines boasts renewable sources like geothermal, biomass, hydro, solar, wind, and biofuels.  Despite solar energy having only a fraction of total energy production, it has steadily gained popularity and preference among electrical system designers. [5]. Renewable energy sources accounted for approximately 6300 MW of installed capacity in 2015. Among them, solar energy has the highest average annual growth rate of 88% from 2014 to 2020 [6]. A solar panel system can power a household with 80% lower carbon emissions than fossil fuels. A study on residential homes in California shows that 0.9% of them have installed solar systems.  However, at least 60% of homes in the area would need to install solar panels to achieve a significant reduction in carbon emissions by 2020. Cost was seen to be the largest barrier; thus, enhanced government subsidies and feed-in tariff, i.e. fixed price of renewable energy set by the government, policies were needed [7]. The approach of combining the use of renewable energy sources and energy efficiency is embedded in the principle of Nearly-Zero Energy Buildings (NZEB) [8, 9]. The European Union defined NZEB as buildings that produce more energy than they consume [10]. In simple terms, an NZEB is a building which has the following characteristics: a) has very low energy use, b) gets power from renewable sources, c) utilize energy-efficient appliances and design, and d) helps reduce electricity consumption and pollution.

Buildings are the major contributors of greenhouse gases, and with NZEB principles, they should achieve reductions of at least 88% - 91% [11]. For Zero-Energy Buildings, energy use is balanced [12]. The energy needs of the building should be considered for this part, and a clear understanding of the energy flow in terms of renewable energy must be considered. Calculations must be made to provide an appropriate renewable energy system to power the required load. Lastly, considerations of the total demand and carbon emissions are also included. If more energy is produced, then there should be clear guidelines on how it is exported [13]. Among other building structures, university buildings can be models for high-value energy retrofitting programs [14]. Public buildings may save about 31% of energy, or 7%, if the investment return is less than three years [15]. Increased energy efficiency reduces the costs of energy services, dependency on energy imports, and negative impacts on the environment [16]. The purpose of retrofitting is to update the existing electrical system of buildings to meet modern safety standards, improve energy efficiency, and incorporate new technology.

In an energy-efficient building design, it is not enough to abruptly reduce the active usage of electrical energy. With this method, the comfort of the dwellers in the building will be compromised, i.e., poor illumination and humid environments. Energy-efficiency efforts also require passive methods like shading and natural lighting. Meanwhile, in terms of cooling loads, shading of the windows from direct sun and other passive measures are the top priority in reducing energy usage for cooling. Certainly, compressor cooling is not within the NZEB energy frames, and other technologies, such as under-floor cooling/heating systems, may be explored [17]. The addition of passive measures is, in fact, contributory to the reduction of energy use. A study in Thailand showed that installing LED lamps and efficient cooling or heating systems is significant in reducing a building's energy demand. The passive measures indicated a payback period of less than six years [18]. Capital investments in a low-carbon future have increased significantly. European countries had identified key components forming the framework of an efficient, low-carbon energy and transport system after 2020. The components included a low-carbon energy source with its supporting systems and infrastructure, requiring an increase in investments estimated to be €270 billion annually [19]. The International Energy Agency (IEA) projects that solar power could generate 22% of the world's electricity by 2050. Developing countries are relevant to this forecast as they have high solar resources, which is becoming a more cost-effective option. There has been an increase of around 40% per annum in the use of such technology. The scenario is coupled with a rapid reduction in the cost of the technology, showing the potential for a large-scale delivery [20]. Several solar power projects have shown that these investments are profitable. One project in Nauru was found to be financially sustainable, given the small share of incremental O&M in NUC's overall expenditure [21]. In the Philippines alone, the Department of Energy has awarded solar power projects listed on its website from Luzon, Visayas, and Mindanao. A potential capacity generation of 23,014.2 MW for Luzon, 2,440.39 MW for Visayas, and 27,032.89 MW for Mindanao is expected from different investors [22]. In 2022, the solar energy producers in the Philippines enjoyed a Feed-in Tariff (FIT) Rate of PhP8.69/kWh. Meanwhile, the reported FIT in 2023 is at Php 9.68 [23]. In computing the return-on-investment, the following are considered: system production, site-specific solar resource, assessment of system cost, direct and indirect capital cost, operation and maintenance, net present value of costs of energy savings [24].

Several studies have been made following the NZEB principle for educational buildings. One study focused on retrofitting a university building in Norway, analysing energy demand, heat consumption, lifecycle environmental impacts, and integration of solar PV systems using lifecycle assessment tools [25]. In Turkey, several energy-efficient scenarios were evaluated in an educational building in Ankara. Simulations were done to test adding PV panels, solar collectors, and wind turbines to achieve NZEB goals [26]. A case study in Japan, focused on a renovated academic building retrofitted into a Zero-Energy or NZEB facility. It also analysed the potential surplus energy of the system [27]. Surplus energy is the excess electricity generated by renewable sources that may be sold back to electrical distribution utilities. Another study for an academic building was retrofitted with an advanced envelope, lighting, and solar PV systems.

Reports showed reductions as much as 30% in energy use and 32% in $CO_2$ emissions. It was also seen to have a short payback period of 2-3 years [28]. One study examined how gymnasium building forms and orientations affect Energy Use Intensity (EUI), Photovoltaic Power Generation (PVPG), and carbon emissions across multiple climate zones. The Building-Integrated Photovoltaics (BIPV) model was evaluated for optimal design configurations [29]. The present study identifies the Tarlac State University (TSU) gymnasium as a candidate for retrofitting into a nearly-zero energy building due to its considerable potential for energy-efficient enhancements. Most of the reviewed literature focused on high-volume, high-occupancy buildings. The on-site survey shows real-world electrical and cooling system limitations of an existing facility, not just simulation models or ideal conditions. It integrates technical retrofitting, economic viability, and NZEB conversion in a Philippine university context, which is underrepresented in current literature.

Named Dr. Mario P. Manese Gymnasium, it has many functions, including hosting indoor sports activities, a venue for physical education classes, commencement exercises, and other school-related activities. It has an open flooring equivalent to a standard basketball court, bleachers, a power and lighting system, air coolers, and a basketball scoreboard. This concept of a multi-use gymnasium led to it becoming a preferred venue for several formal assemblies, like civic and religious assemblies, and other local community events.

Considered as large-volume and high-occupancy public buildings, gymnasiums should be conducive for sports events. According to a Japanese sportswear producer, sports and climate change are highly related [30]. There were reports of cancelled outdoor sports events and players hospitalized due to intense rain and harsh heat waves [31]. The SEA Games, for instance, were part of the Race to Zero global campaign. The temperature as high as 40 degrees Celsius during the opening of the 2023 Southeast Asian Games in Cambodia caused worry among athletes and sports officials about the disruption of events. The Cambodian sports officials, however, assured all delegates that green initiatives to counter the effects of climate change during the SEA games were in place because the environmental issue was also a major concern for the Cambodian athletes [32].

At present, the condition of the TSU gymnasium is deemed not conducive to longer periods of usage due to perceived heat. This scenario is being felt during Intramurals, which occur in April as both athletes and spectators flock to the Gymnasium. If additional air-conditioning units are installed, an upgraded electrical facility must also be provided. With this, the Tarlac State University can have a role to play in limiting the rise in global temperature by transforming the Dr. Mario P. Manese Gymnasium into a nearly zero-carbon-emission gymnasium. It endeavours to align with the principles of zero carbon emissions. Furthermore, it investigated certain issues and challenges regarding the integration of the solar PV system into the existing electrical service of the Gymnasium. Finally, the present study sets out to realize the following specific objectives: a) to describe the existing state of the TSU Gymnasium's electrical infrastructure, b) to implement improvements that result in a facility capable of generating more energy than it consumes by optimizing lighting and cooling systems to enhance energy efficiency by considering the integration of an on-grid solar PV system to contribute to sustainable energy generation, and c) to perform solar electric investment analysis to look into the viability of the said project.

## 2. Materials and Methods

Both quantitative and qualitative techniques were used to gain a better understanding of developing a design for the TSU Gymnasium, leading to making it a nearly zero-energy building. The sequence starts with employing rigorous quantitative assessment of the existing state of the Gymnasium's electrical infrastructure to implement improvements aligned with sustainable energy generation and reducing environmental impact. Finally, a qualitative method was used to elaborate on the quantitative results. The Dr. Mario P. Manese Gymnasium (TSU Gymnasium) is located at the Main Campus of Tarlac State University, situated along Romulo Boulevard, San Vicente, Tarlac City. The facility is being used both during graduation rites, mini concerts, and other culminating activities. The facility is found to be at 15.8 °N latitude and 120.59 °E longitude, while the elevation of the place is 51 m [33]. These could be significant in utilizing the PVSyst software in the determination of the appropriate solar PV system components.

### 2.1. Data Gathering Tools
To help the researchers achieve the objectives, the following were utilized to gather data. The said instruments were used independently or in combination to gain a deeper understanding of the study:

1. Document Review - Blueprints are especially necessary for the study to be conducted. The researchers investigated records of the as-built floor plan of the Gymnasium as well as the energy consumption of the Gymnasium. Permission from the administrators was obtained to acquire the needed documents.
2. Interview - The Gymnasium is an old structure, so tenured employees at the University helped provide insights and valuable knowledge about the Gymnasium.
3. Site Visit - Walk-through inspections were beneficial to the researchers during the assessment and the design stage. Actual measurements and observations had become inputs on how lighting and electrical equipment are located on the plan.
4. Simulations from PVSyst Software - The said tool was used to project outputs, considering the design and investment cost analysis, as well as for the computation of carbon emissions. In using PVsyst, the designer has an array of tools within the software that suggest an array/system configuration, which allows preliminary simulation. It also includes a color-coded warning/error messaging system. It also displays any issues in the design if there is a mismatch in the parameters [34]. The University of Geneva developed the tool, which features modelling, sizing, simulating, and analyzing PV systems. The Users can generate results that can bring crucial insights into engineering aspects [35].
5. Analysis from Psychrometric Analyzer - Computations for the appropriate cooling system considered the sensible heat and volume flow rate required for the Gymnasium. Equation 1 shows the sensible heat formula to calculate the required flow rate, assuming 70 watts of sensible heat per person from 1500 occupants.

$$Q_s = (V/v) C_p (T_2 - T_1) \qquad (1)$$

Where:

$Q_s$ is the specific heat transfer
$v$ is the specific volume
V is the volume flow rate
$C_p$ Is the specific heat at constant pressure?
$(T_2 - T_1)$ Is it the change in temperature?

### 2.2. Data Gathering Procedures

Permission from the administrators of TSU was obtained by writing a letter to gather data with regard to the TSU Gymnasium. An on-site survey of the Gymnasium was employed by taking note of the number and rating of lighting, air-conditioning units, and convenience outlets. The researchers also determined the room data, particularly the dimensions and surface reflectance of the Gymnasium. Cavity data represents the room, ceiling, and floor cavity. The fixture data, such as the manufacturer, model, lumens/lamp, etc, were also noted. Based on the data gathered, the researchers computed the appropriate number and rating of lighting fixtures for the current needs of the Gymnasium, following standard electrical lighting design principles. For air-conditioning units, the researchers used an online Psychrometric Analyzer to compute the flow rate for 1500 occupants. The computed value was then used as input to compute the number of air conditioning units. The load for convenience or power outlets was estimated at 8 kVA per square meter. After determining the total electrical load, solar panel sizes and investment analysis were performed using PVSyst software. The simulation included system design and production, system cost, operation and maintenance, financial factors, and the value of electricity generated by the system. Carbon footprints were also included in the report.

A copy of the as-built plan of the TSU Gymnasium became the basis for the design of the PV System, while the energy consumption of the Gymnasium for the last 3 years was used in the computation of the financial analysis. The straightforward, simple payback formula is a preferred evaluation metric for solar installers. The calculation does not consider several real variables, such as time value of money, energy escalation rates, rate structure, and opportunity costs. Data were analyzed by a combination of quantitative and qualitative approaches. Interpretation and analysis of figures and tables were done to show significant findings. Meanwhile, quantitative analyses were utilized by the researchers with the aid of PVSyst Software 7.3.1 and Psychrometric Analyzer 6.8, considering some estimated parameters.

## 3. Results and Discussion
### 3.1. Electrical Infrastructure of TSU Gymnasium
#### 3.1.1. Lighting Loads

The basketball court of the Gymnasium utilizes high bay metal halide lamps to light the basketball court. The lamps are a popular choice for stadiums, sports fields, parking lots, and street lighting in urban areas, and they were first introduced in the 1960s [36]. They belong to High-Intensity Discharge (HID) lamps, which are perfect for wide areas like roads and gymnasiums [37]. Photos in Figure 1 show some of the lighting installed in the Gymnasium. However, the end of this type of lamp is looming due to efforts to reduce carbon footprint. Their high maintenance, long warm-up, high heat production, occasional explosions and due to mercury content [38]. Meanwhile, several fluorescent lamps and Compact Fluorescent Light (CFL) bulbs are used to light the stage and dressing rooms backstage. These lamps are considered a smarter choice over the obsolete incandescent bulb [39]. However, since the color of these types of lamps is not meant for creative stage lighting, organizers compensate for the issue by renting additional lighting during events.

#### 3.1.2. Air-Conditioning Loads

There is a total of twelve air-coolers, two of which are movable, and ten are installed in the windows used in the Gymnasium. Additionally, eight exhaust fans and fifteen electric fans are positioned around the court, stage, and below the bleachers during events like graduation rites, concerts, and other culminating activities. Some of the photos of the said equipment are shown in Figure 2. The current cooling arrangement-comprising multiple window-mounted air-coolers, exhaust fans, and electric fans-reveals a fragmented and inefficient approach to indoor thermal management in the Gymnasium.

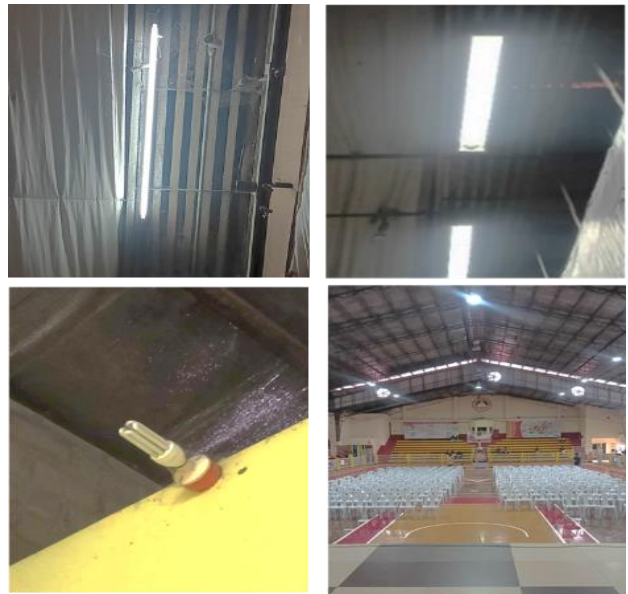

**Fig. 1 Gymnasium lighting loads**

This system is unlikely to provide consistent or adequate cooling for large-scale events, contributing to excessive energy use without achieving optimal comfort. These findings underscore the necessity for a centralized, energy-efficient

HVAC solution aligned with NZEB design principles to ensure sustainable cooling and improve occupant comfort during high-occupancy events.

*3.1.3. Convenience or Power Outlets*

Currently, the Gymnasium has two convenience or power outlets in the dressing rooms at the back of the stage, two on the stage, and eight in the court area. There is a control room for the sound system, but it is not being used anymore, according to the personnel assigned to the facility.

The current configuration of electrical outlets in the Gymnasium is insufficient to meet the demands of modern, high-occupancy, and multimedia-driven events. The limited number and poor distribution of outlets restrict functionality, increase reliance on temporary solutions, and may present safety concerns. Furthermore, the underutilization of the control room reflects a broader issue of outdated infrastructure. These findings imply a critical need to modernize the electrical system-to support current usage and to enable the integration of energy-efficient technologies and renewable systems consistent with the transformation to a Nearly Zero-Energy Building (NZEB).

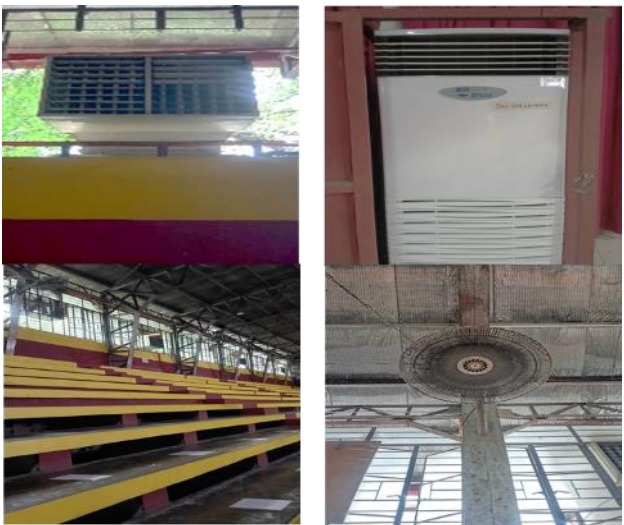

**Fig. 2 Gymnasium's air-cooling units**

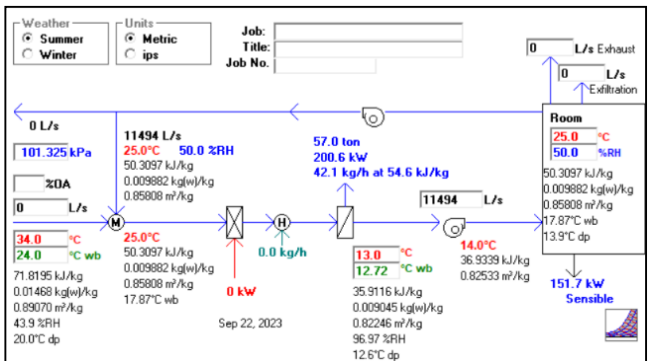

**Fig. 3 Data run using psychrometric analyzer**

### 3.2. Improvements that Resulted in a Facility Capable of Generating More Energy than it Consumes
*3.2.1. Lighting System*

The results in Table 1 show the lighting fixture data of the area of the Gymnasium where big events take place, including the bleachers, with the energy-efficient lighting system.

Accordingly, replacing high bay halide lamps with LED high bay fixtures is the most effective way to increase the Gymnasium's energy efficiency. This can potentially cut lighting energy costs, with energy savings of 68% for the payback period of 4.92 years [40].

**Table 1. Fixture data for the court area of the TSU gymnasium**

| A. Room Data | | |
|---|---|---|
| Room Dimensions | Length | 31 m. |
| | Width | 19.51 m |
| | Floor Area | 604.81 m² |
| | Ceiling Height | 10.6 m |
| Surface Reflectance | Ceiling | 80 |
| | Wall | 50 |
| | Floor | 35 |
| | Fixture Mounting Height | 7.9 m |
| **B. Cavity Data** | | |
| Room Cavity | Height | 7.9 m |
| | Ratio | 3 |
| Ceiling Cavity | Height | 2.7 m |
| | Ratio | 1.2 |
| | Effective Reflectance | 0.64 |
| Floor Cavity | Height | 0 m |
| | Ratio | 0 |
| | Effective Reflectance | 0.34 |
| **C. Fixture Data** | | |
| Manufacturer | | FELCO |
| Model & Catalog No. | | FHB5200DL |
| Lamps/Fixture | | 1 |
| Lumens/Lamp | | 26000 |
| Coeff. Of Utilization | | 0.805 |
| LDD Factor | | 0.88 |
| LLD Factor | | 0.7 |
| **D. Number of Fixtures** | | |
| Desired Lighting Level | | 500 lux |
| No. of Fixtures | | 24 |

*3.2.2. Energy Efficient Cooling System*

A thorough analysis was conducted to determine an energy-efficient cooling system for the gymnasium. By using Equation 1 mentioned above, the required flow rate was calculated as 11.494 litres/second. Having established the flow, the Psychrometric Analyzer (Figure 3) was used to obtain data for calculating the number of Air Conditioning

Units (ACU). The result, as shown below, indicated a requirement of 18 floor-mounted units, each with a capacity of 3.0 TR and a power demand of 4.5 kVA.

$$No. of\ ACU = \frac{(200.6 \ kJ/s)(3600s/hr)}{40,090\ kJ/hr} \approx 18\ units$$

In the determination of the total load provision for the entire system, the load for convenience outlets was also estimated based on the general provision of 8 Volts-Ampere per square meter. The Gymnasium has a total area of 1,608 m$^2$; hence, the expected load is 12.86 KVA. As summarized in Table 2, the total estimated load for the TSU Gymnasium, considering LED and inverter air conditioning units, is 98.66 KVA.

Meanwhile, the summary for the improvements mentioned above can be seen in Table 3. The researchers used the PVsyst software to formulate the on-grid solar PV system design for the TSU Gymnasium. This simulation tool used the 2020 meteorological database.

**Table 2. Estimated load for the gymnasium**

| Load | Va Per Unit | Units | Subtotal |
|---|---|---|---|
| Lightings | 200 VA | 24 | 4.8 KVA |
| ACUs | 4500 VA | 18 | 81 KVA |
| Outlets | - | - | 12.86 KVA |
| Total | | | 98.66 KVA |

**Table 3. Improvements in no. of lightings, ACUs, and convenience or power outlets in the TSU Gym**

| Load | Before | After |
|---|---|---|
| Lightings | 10 Fluorescent lamps 10 metal halide lamps | 24 High Bay LED lamps |
| ACUs | 12 air-coolers, 8 exhaust fans 15 electric fans | 18 - 3 TR floor-mounted inverter units |
| Outlets | 10 convenience outlets | 72 convenience outlets |

The analysis indicated that the best configuration for the Gymnasium involved the installation of 252 units of PV arrays and 2 units of inverters to meet its energy requirements sustainably. Assumptions for the simulation included no shading and no 3D scene defined. There are two fixed plane orientations with tilts/azimuths of 15/-8º and 15/172º.

*3.2.3. Solar PV System*

Table 5 shows the manufacturer, model, and other characteristics pertaining to the PV modules, while Figure 4 shows the single-line diagram of the system. Considering the use of LG Electronics products, the system needs 252 units of modules comprising 6 strings, 21 in series. Meanwhile, Tables 4 and 5 show a detailed list of characteristics and other module design information of the system.

*3.3. Solar Electric Investment Analysis*

The characteristics of the PV System are summarized in Figure 5. The administration of TSU needs a total of about Php 6.9 million to put up the system.

**Table 4. System PV power/inverter characteristics**

| Total PV Power | |
|---|---|
| Nominal (STC) | 100 kWp |
| Total | 252 modules |
| Module area | 457m$^2$ |
| Cell area | 415 m$^2$ |
| Inverter | |
| Manufacturer | SofarSolar |
| Model | SOFAR 40000 TL |
| Unit nominal power | 40.00 KWac |
| Number of inverters | 2 units |
| Total power | 80.0 KWac |
| Operating Voltage | 250-960 V |
| Pnom ratio (DC:AC) | 1.24 |

**Table 5. Module design for the 100 kWp PV system**

| Parameter | Specification |
|---|---|
| Manufacturer | LG Electronics |
| Model | LG 395 Q1C-A6 |
| Unit nominal power | 395 Wp |
| Number of PV modules | 252 units |
| Nominal (STC) | 100 kWp |
| Modules | 6 strings; 21 in series |

The total cost includes the PV modules, inverters, accessories, and other components, as well as installation. The figure also shows other details for the unit costs of components such as wiring, accessories, and installation costs per module/inverter. Surplus energy may be priced at 62 MWh/year, which can generate Php 4.614 per kWh.

Meanwhile, as shown in Figure 6, the operating costs account for Php 305,275.40 per year. It comprises replacement for inverters, repairs, and maintenance.

The system is assumed to be operable for 20 years and to be implemented in the year 2024. Other assumptions were neglected, like income variation over time and income-dependent expenses.

Meanwhile, the straight-line method was considered for depreciation. As mentioned in the introduction, the approved feed-in tariff for solar power generation in the Philippines is at Php 9.68 per kWh. The summary of the financial analysis and assumptions set by the researchers in the simulation is shown in Figure 7.

The return-on-investment for the installation of the system could be covered in the first 9 years of operation, as shown in Figure 8.

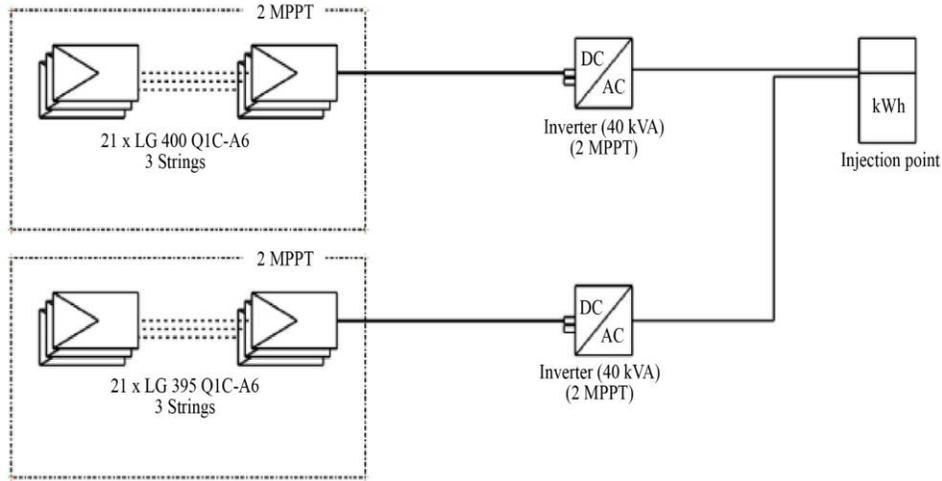

**Fig. 4 PV System single line diagram for the TSU gymnasium**

| Item | Quantity units | Cost PHP | Total PHP |
|---|---|---|---|
| PV modules | | | |
|   LG 400 Q1C-A6 | 126 | 23,278.00 | 2,933,028.00 |
|   LG 395 Q1C-A6 | 126 | 20,278.00 | 2,555,028.00 |
|   Supports for modules | 252 | 1,000.00 | 252,000.00 |
| Inverters | | | |
|   SOFAR 40000TL | 2 | 263,188.50 | 526,377.00 |
| Other components | | | |
|   Accessories, fasteners | 1 | 50,000.00 | 50,000.00 |
|   Wiring | 1 | 50,000.00 | 50,000.00 |
|   Combiner box | 2 | 25,000.00 | 50,000.00 |
|   Surge arrester | 4 | 12,500.00 | 50,000.00 |
| Installation | | | |
|   Global installation cost per module | 252 | 1,000.00 | 252,000.00 |
|   Global installation cost per inverter | 2 | 25,000.00 | 50,000.00 |
|   Transport | 1 | 50,000.00 | 50,000.00 |
|   Settings | 1 | 25,000.00 | 25,000.00 |
|   Grid connection | 1 | 50,000.00 | 50,000.00 |
| | | Total | 6,893,433.00 |
| | | Depreciable asset | 6,316,433.00 |

**Fig. 5 Installation cost of the PV system in the TSU gymnasium**

**Operating costs**

| Item | Total PHP/year |
|---|---|
| Maintenance | |
|   Provision for inverter replacement | 105,275.40 |
|   Repairs | 100,000.00 |
|   Cleaning | 100,000.00 |
| **Total (OPEX)** | **305,275.40** |

**System summary**

| | |
|---|---|
| Total installation cost | 6,893,433.00 PHP |
| Operating costs | 305,275.40 PHP/year |
| Unused energy | 78.9 MWh/year |
| Energy sold to the grid | 62.0 MWh/year |
| Cost of produced energy (LCOE) | 4.614 PHP/kWh |

**Fig. 6 Projected operating costs of the PV system of the TSU gymnasium**

At this rate, the TSU administration is expected to gain a cumulative profit starting in the year 2034, approximately Php 8.3 million for the entire lifespan of the project. To attain the payback period in 10 years, the system performance ratio is expected to be at 85.2%. The investment for the PV System for the TSU Gymnasium is within the average payback period of 7-10 years for solar power [41]. However, the administration must take into account other factors, such as electrical rewiring, structural reinforcements, labor costs, permit and compliance costs. Details for the economic evaluation are shown below in Figure 9. Ultimately, the effort in designing the TSU Gymnasium to be within the bounds of the NZEB framework is to contribute to the reduction of carbon emissions. According to the results of the software shown in Figure 10, a total of 2028.3 tons of Carbon Dioxide (tCO2) would be prevented from being released into the environment if the designed PV system for the TSU Gymnasium is implemented.

**Fig. 7 Financial analysis report**

**Fig. 8 Detailed economic results of the TSU gymnasium PV system**

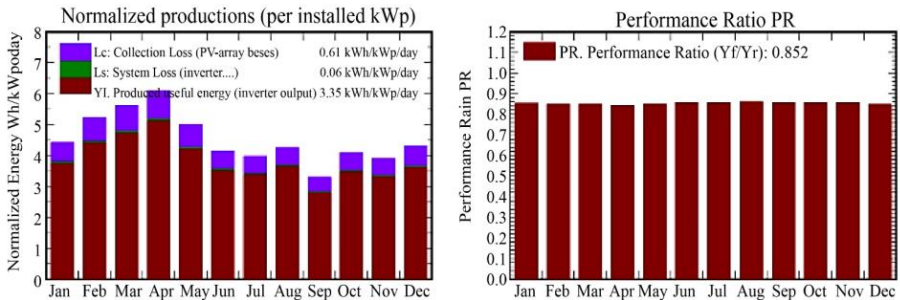

**Fig. 9 Economic evaluation of the PV system of the TSU gymnasium**

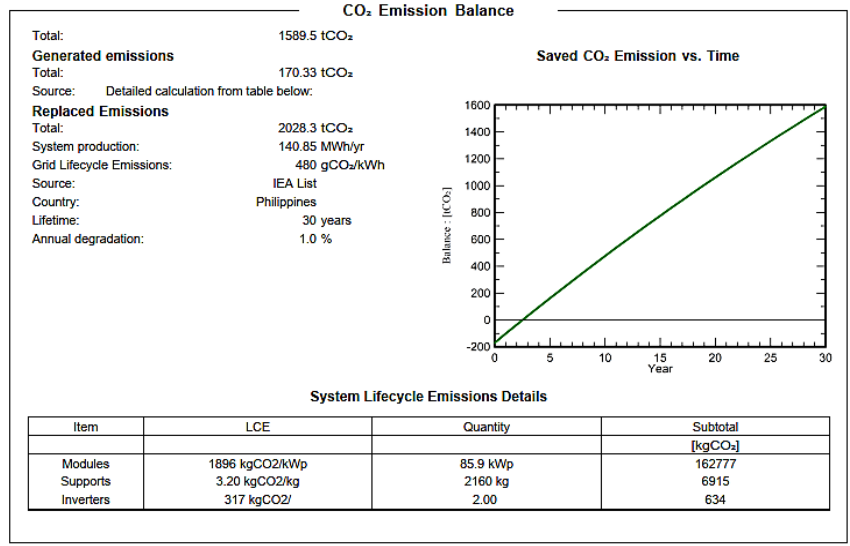

**Fig. 10 Economic evaluation of the PV system of the TSU gymnasium**

## 4. Conclusion

The electrical infrastructure of the TSU Gymnasium is outdated, the lighting is not energy efficient, and there are too few convenience or power outlets. Additionally, there is insufficient cooling equipment to maintain a comfortable temperature inside the Gymnasium. The installation of the PV system for the TSU Gymnasium, combined with the implementation of efficient energy lighting and cooling systems, is a feasible solution for implementing improvements that result in a facility capable of generating more energy than it consumes. The results of this study show that the payback period is within the average range, making it a cost-effective investment for the University. Aside from the cost of the PV System, adherence to engineering design standards will mean additional costs to replace the metal halides with LED high bay lamps, installation of additional air conditioning units, and provision of additional convenience outlets. These additional costs should be considered when evaluating the feasibility of the project. If TSU implements a Gymnasium into a nearly zero-energy facility, it will help reduce carbon footprints. The researchers recommend that the integrity of the existing roofing system of the TSU Gymnasium be analyzed. The total cost of putting up the whole electrical system, including new lighting, cooling, and convenience loads, must be properly calculated to determine the total cost of implementing the whole NZEB project. Other factors in the economic evaluation should also be considered by the TSU administration to determine a more stringent result. Net metering should be applied for by the TSU administration from the local distribution utility.

## Acknowledgments

The authors would like to express their gratitude to the personnel in charge of the TSU Gymnasium for their assistance during the survey.